\documentclass[twocolumn,prb,showpacs,preprintnumbers,amsmath,amssymb]{revtex4}

\usepackage{subfigure}
\usepackage{graphicx}
\usepackage{dcolumn}
\usepackage{bm}
\usepackage{setspace}

\begin{document}


\title{Signature of  pseudogap formation in the density of states of
underdoped cuprates
} 

\author{A. J. H. Borne$^{1,2,3}$}
\author{J. P. Carbotte$^{4,5}$}
\author{E. J. Nicol$^{1,2}$}%
\email{enicol@uoguelph.ca}
\affiliation{$^1$Department of Physics, University of Guelph,
Guelph, Ontario, Canada N1G 2W1} 
\affiliation{$^2$Guelph-Waterloo Physics Institute,
University of Guelph, Guelph, Ontario, Canada N1G 2W1}
\affiliation{$^3$PHELMA Grenoble INP, Minatec, 3 Parvis Louis N\'eel,
BP 275, 38016, Grenoble, Cedex 1, France}
\affiliation{$^4$Department of Physics and Astronomy, McMaster
University, Hamilton, Ontario, Canada L8S 4M1}
\affiliation{$^5$The Canadian
Institute for Advanced Research, Toronto, Ontario, Canada M5G 1Z8}

\date{\today}

\begin{abstract}
The resonating valence bond spin liquid model for the underdoped
cuprates has as an essential element, the
emergence of a pseudogap.
 This new energy scale
introduces asymmetry in 
the quasiparticle density of states because it is associated with 
the antiferromagnetic
Brillouin zone. By contrast, superconductivity develops on the Fermi
surface and this largely restores the particle-hole symmetry 
for energies below
the superconducting energy gap scale. 
In the highly underdoped regime, these two scales can be separately
identified in the density of states and also
partial density of states  for each
fixed
angle in the Brillouin zone. From the total density of
states, we find that the pseudogap energy scale manifests itself differently
as a function of doping for positive and negative bias. Furthermore,
we find evidence from recent scanning tunneling spectroscopy
data for asymmetry in the positive and negative bias of the extracted
$\Delta(\theta)$ which is in qualitative agreement with this model. 
Likewise, 
the slope of the linear low energy density of states
is nearly constant in the underdoped regime while it increases significantly
with overdoping in agreement with the data.

\end{abstract}


\maketitle

\section{Introduction}

BCS theory modified to account for $d$-wave symmetry of the superconducting
order parameter has provided a solid basis for a first understanding of the
properties of the cuprates around optimum and in the overdoped regime. However,
the underdoped region of the phase diagram provides challenges to such a 
simple approach. These systems present many features that have been considered
anomalous and are not part of simple BCS theory. While the subject
remains controversial and very different ideas have been put forward to 
understand these anomalies, a recently developed model by Yang, Rice and 
Zhang (YRZ)\cite{yrz} which is based on a spin liquid resonating valence bond 
approach, has had considerable success in this direction. It is very different
from the preformed pair model\cite{emery} where a superconducting gap forms
at a high temperature, the pseudogap temperature $T^*$,
 and the superconductivity
appears only at a lower temperature $T_c$ as a result of the onset of phase 
coherence. The YRZ model has a central element, a new energy scale, the
pseudogap $\Delta_{\rm pg}$, which is responsible for changes in the
electronic structure of the normal state above $T_c$. The pseudogap is distinct from the superconducting gap 
$\Delta_{\rm sc}$ and the superconducting state is conceived as forming from
this new normal pseudogap phase which is quite different from an ordinary
Fermi liquid state. In the YRZ model, the large Fermi surface of Fermi liquid
theory (FLT) reconstructs into hole and electron pockets as a result of the
growth in pseudogap at doping levels $x$, in the underdoped region of
the cuprate phase diagram, with $x<x_{c}$ where $x_{c}$ is the doping 
associated with a quantum critical point (QCP).

The model is related but different from other competing order proposals
such as D-density wave formation\cite{laughlin,zhu} and
has the desirable property that, in its final form, it remains simple and
has been successfully applied to the calculation of many superconducting
properties\cite{belen,jamescv,jamesarpes,emilia,yrzarpes,bascones,yrzandreev,kent}.
Besides a self-energy which accounts for the formation of a pseudogap on the
antiferromagnetic Brillouin zone (AFBZ) boundary, the model has
Gutzwiller factors modifying the underlying band structure parameters.
These narrow the bands as correlation effects become more important.
Also, a Gutzwiller factor accounts for the loss of coherence which greatly
reduces the weight of the remaining quasiparticle peak as the incoherent
background increases. These elements account for the approach to the
Mott insulating state which is best understood near half-filling in
a localized picture of electron dynamics. The hopping from
one site to another is blocked by a large Hubbard $U$ which describes the energy
cost for double occupancy. The pseudogap and 
AFBZ then plays a role similar to a
band gap at the Brillouin zone in ordinary band theory but with essential
differences. For example, in YRZ, the bands are not filled rigidly with
decreasing doping but instead undergo profound changes as the pseudogap increases.

Among the properties already calculated  and compared with experiment
are Raman\cite{belen,jamesarpes}, specific heat\cite{jamescv},
optical properties\cite{emilia}, aspects of angular resolved photoemission
spectroscopy (ARPES) and scanning tunneling spectroscopy (STS)\cite{yrzarpes}
including the checkerboard pattern\cite{bascones}, Andreev tunneling\cite{yrzandreev}
 and also the penetration
depth\cite{kent}. Each of these properties show behaviors which cannot be 
understood in $d$-wave BCS nor in it extensions to include inelastic scattering\cite{nicol,ewald,inversion,tu,hwang,marsiglio,nicolmfl,basov},
anisotropy\cite{leung1,donovan2,donovan3,mac,mitrovic}, or strong coupling effects rooted in 
Eliashberg theory\cite{mac,mitrovic,daams}. Among the previously considered
anomalous properties that are now understood are the two distinct
gap scales seen in Raman spectra. The B$_{2g}$ peak decreases with
decreasing doping while the B$_{1g}$ scale increases instead\cite{belen,jamesarpes}.
The normalized jump in the specific heat drops rather precipitously as $x$
decreases towards the bottom of the superconducting dome\cite{jamescv}.
New structures are seen in the optical self-energy and the two energy
scales found in the partial optical sum as a function of energy are understood\cite{emilia}.
The rapid drop in the superfluid density at zero temperature with decreasing
$x$ while the slope of the low temperature linear in $T$ law is relatively
only weakly changed is also shown\cite{kent}. 
Encouraged by these successes, here we consider
the quasiparticle density of states (DOS).

In Sec. II, we present the formalism of YRZ\cite{yrz}
for the electron spectral density in the underdoped regime. It includes
pseudogap formation below a QCP at a doping
of $x=x_c=0.2$. We describe how this new energy scale modifies the electronic
structure from the usual large Fermi surface of FLT
to Luttinger pockets. There are two energy branches in the theory $E^\pm_{\boldsymbol{k}}$,
with $E^-_{\boldsymbol{k}}$ giving a hole pocket and $E^+_{\boldsymbol{k}}$ an electron
pocket, the latter only in a restricted doping range
just below $x_c$. For very underdoped samples, only the hole pocket remains.
In Sec. III, we present our results for the quasiparticle DOS with and without
superconductivity and also break up the results into their partial contributions from each of the two energy
branches separately. We discuss how the pseudogap alone introduces an asymmetry
between positive and negative biases in the DOS $N(\omega)$ and how
superconductivity overrides this effect and so restores particle-hole symmetry
at small energies of order the superconducting gap energy $\Delta^0_{\rm sc}$.
At higher energy, asymmetry remains. We also show that modifications in the 
DOS introduced by superconductivity which are confined to a range of a few
times $\Delta^0_{\rm sc}$ in the case of a Fermi liquid extend instead
over the range of the pseudogap energy scale in YRZ. In Sec. IV, we consider
decomposing the total quasiparticle DOS into partial contributions from
fixed angle $\theta$ measured from $(\pi,\pi)$ in the upper right quadrant
of the Brillouin zone. These partial distributions 
can display several structures. Nevertheless, upon examination we are able to
define for each direction $\theta$ an energy gap associated with a specific peak
in the partial DOS. This peak which is taken mainly, but not always, 
to correspond to the smallest energy
closest to $\omega=0$, is sometimes the superconducting gap peak
but can also be a pseudogap peak. The energies extracted in this way
show considerable anisotropy between positive and negative biases which is a fundamental
characteristic of the model used here. We find evidence in experiment for
this anisotropy and provide a comparison with STS
data.
In Sec.~V, we discuss the limit
of low bias where the DOS is linear in $\omega$. We recover the FL result
with an extra Gutzwiller factor reflecting the strong correlations in the
system.
Section VI contains a summary and conclusions.

\section{Formalism}

The spectral function for the coherent part of the electronic propagator
in the model of Yang, Rice and Zhang\cite{yrz} takes the form

\begin{equation}
A(\boldsymbol{k},\omega)=\sum_{\alpha=\pm}g_t(x)W_{\boldsymbol{k}}^\alpha
[(u_{\boldsymbol{k}}^\alpha)^2\delta(\omega-E^\alpha_S)+(v_{\boldsymbol{k}}^\alpha)^2\delta(\omega+E^\alpha_S)],
\label{eq:A}
\end{equation}
where  $E^{\alpha}_S=\sqrt{(E_{\boldsymbol{k}}^{ \alpha })^2  +
  \Delta _{\rm sc}^2(\boldsymbol{k})}$ are the quasiparticle energies in the superconducting state
with $(u_{\boldsymbol{k}}^\alpha)^2$ and $(v_{\boldsymbol{k}}^\alpha)^2$
the corresponding Bogoliubov amplitudes. In Eq.~(\ref{eq:A}), the
$W_{\boldsymbol{k}}^\alpha$ ($\alpha=\pm$) are the weighting factors of the YRZ
theory which involve the input pseudogap $\Delta_{\rm pg}(\boldsymbol{k})$
and do not change in the superconducting state. They depend on the electronic
band structure energies $\xi_{\boldsymbol{k}}$ as well as the Umklapp
surface energy $\xi^0_{\boldsymbol{k}}$. Specifically,
\begin{equation}
W_{\boldsymbol{k}}^ \pm   = \frac{1}{2}\left( {1 \pm
  \frac{{\tilde{\xi_{\boldsymbol{k}}}  }}{E_{\boldsymbol{k}}}}
\right).
\label{eq:W}
\end{equation}
 with 
$\tilde{\xi_{\boldsymbol{k}}}  = (\xi_{\boldsymbol{k}}  + \xi_{\boldsymbol{k}}^0 )/2$ and
$E_{\boldsymbol{k}} = \sqrt {\tilde{\xi_{\boldsymbol{k}}}^2  + \Delta
  _{\rm pg}^2(\boldsymbol{k})}$.
In the nonsuperconducting state, the energies have two branches
$E_{\boldsymbol{k}}^ \pm  = ({\xi_{\boldsymbol{k}}  - \xi_{\boldsymbol{k}}^0 })/2 \pm E_{\boldsymbol{k}}$
and in the superconducting state, the Bogoliubov weights are given in terms of these by
\begin{eqnarray}
(u^\alpha_{\boldsymbol{k}})^2&=&\frac{1}{2}\biggl(1+\frac{E_{\boldsymbol{k}}^\alpha}{E^\alpha_S}\biggr),\\
(v^\alpha_{\boldsymbol{k}})^2&=&\frac{1}{2}\biggl(1-\frac{E_{\boldsymbol{k}}^\alpha}{E^\alpha_S}\biggr).
\end{eqnarray}
In the YRZ paper, the band energies, taken to include up to third nearest
neighbor hopping, are
$\xi_{\boldsymbol{k}} = - 2t(x)(\cos
k_xa  + \cos k_ya ) - 4t^{\prime}(x) \cos k_xa \cos k_ya - 2t''(x)(\cos
2k_xa  + \cos 2k_ya )-\mu_p$,
where $\mu_p$ is the chemical potential adjusted to obtain the correct
number of electrons, through a procedure based on Luttinger's theorem.
 The Umklapp surface energy is where  
$\xi_{\boldsymbol{k}}^0  =  - 2t(x)(\cos k_xa  +
\cos k_ya)$ equals zero. Here $a$ is the two-dimensional CuO$_2$ plane
lattice parameter. The form of the 
hopping coefficients $t(x)$, $t^{\prime
}(x)$, and $t^{\prime\prime
}(x)$ are fixed in the YRZ model\cite{yrz}
 and will not be changed in this work except
to note that $t_0$ enters as a proportionality factor in these
hoppings and consequently
all our results scale by $t_0$ and so this parameter can be varied at will.
The Gutzwiller factor $g_t(x)$ which appears as a simple multiplicative
factor in Eq.~(\ref{eq:A}) provides a measure of the remaining quasiparticle
strength in the coherent part of the Green's function. Along with a second
Gutzwiller factor $g_s(x)$, 
it also enters the band structure, through $t(x), t'(x)$ and $t''(x)$,
 which provides narrower
bands as the  doping $x$ is reduced towards the Mott insulating state at half 
filling. Specifically, $g_t(x)=2x/(1+x)$ and $g_s(x)=4/(1+x)^2$.

For the input superconducting gap, YRZ take 
\begin{equation}
\Delta_{\rm sc}(\boldsymbol{k})=\frac{\Delta_{\rm sc}^{0}(x)}{2}(\cos
k_xa -\cos k_ya)
\label{eq:gapk}
\end{equation}
and likewise the same form for $\Delta_{\rm pg}(\boldsymbol{k})$ with the
gap amplitude $\Delta^0_{\rm pg}(x)$ replacing the superconducting gap
amplitude in Eq.~(\ref{eq:gapk}).
Both have the simplest $d$-wave form characterized by the lowest
harmonic having the required symmetry. The amplitudes in Eq.~(\ref{eq:gapk})
and equivalently for the pseudogap are
\begin{eqnarray}
\Delta^{0}_{\rm
    sc}(x)&=&0.14t_0[1-82.6(x-0.2)^2]\\
\Delta^{0}_{\rm
    pg}(x)&=&3t_0(0.2-x), 
\end{eqnarray}
where both are taken to be proportional to $t_0$ and the QCP at which 
the pseudogap gap becomes nonzero is 
$x_c=0.2$, where the superconducting gap is also taken to have its optimum
value. This last condition can easily be relaxed to have the maximum gap
at 0.16 instead (as in experiment).

\begin{figure}
\includegraphics[width=0.48\textwidth]{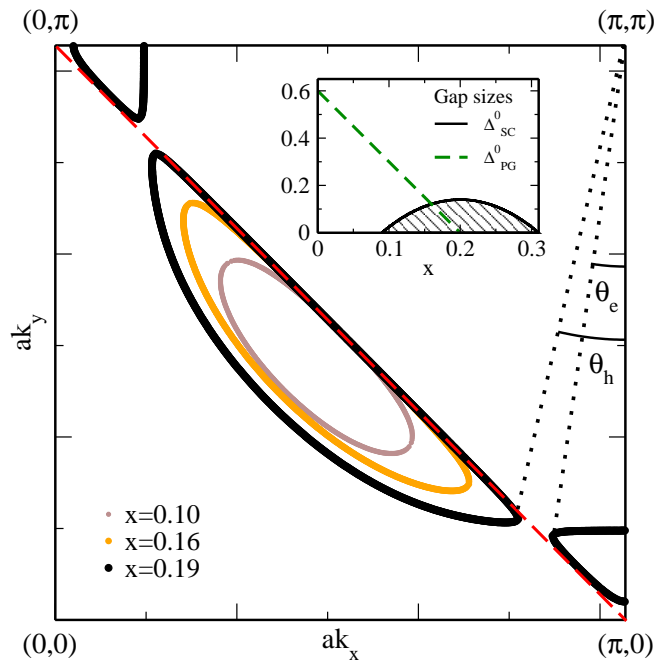}
\caption{\label{fig1} (Color online) The reconstructed Fermi surface contours 
for three values of doping: $x=0.10$ (brown), 0.16 (yellow), and 0.19 (black),
shown in the upper right quadrant of the square Brillouin zone.
The black contours have both hole and electron Luttinger pockets
(located around the nodal and antinodal directions, respectively).
The angles measured from $(\pi,\pi)$ which define the end of these pockets 
are $\theta_h$ and $\theta_e$, respectively. The inset shows the superconducting
dome and pseudogap line as a function of doping defined
via $\Delta^0_{\rm sc}$ and $\Delta^0_{\rm pg}$ in the YRZ model.
}
\end{figure}

In Fig.~\ref{fig1},
 we show how the Fermi surface contours evolve with doping. At the
QCP ($x=0.2$) there is no pseudogap in the YRZ model and the Fermi surface  is the
usual large open surface of Fermi liquid theory (FLT). As $x$ is lowered into the underdoped
regime, the Fermi surface reconstructs and Luttinger contours of zero energy emerge.
They correspond to either $E^-_{\boldsymbol{k}}=0$ or $E^+_{\boldsymbol{k}}=0$.
The solutions for the first case give the hole pockets centered on the nodal
direction and these exist for $\theta=\pi/4$ to $\theta_h$ as indicated in the figure.
For the case of $x=0.19$ but not for the other values of doping shown, there
is an additional electron pocket ($E^+_{\boldsymbol{k}}=0$) located in the
region between the AFBZ boundary (dashed red line) and the Brillouin zone
which extends from $\theta=\theta_e$ to $\theta=0$, where $\theta$ is an
angle measured from the origin $(\pi,\pi)$ in the right hand upper quadrant
of the Brillouin zone as shown. In both cases for hole and electron contours,
solutions to the equation $E^\pm_{\boldsymbol{k}}=0$ 
when they exist (for a given angle $\theta$) always come in pairs.
For some angles however, there is no solution at all. When two solutions
exist, the backside of the Luttinger pocket closest to the AFBZ boundary
(red dashed line) has only a small weight $W^\pm_{\boldsymbol{k}}$ 
as compared with the front part which is orientated towards the center of the Brillouin
zone for the hole pocket and oppositely for the electron pocket.
Both of these have weight of order one and this is the piece of the Fermi
surface (FS) which corresponds to the Fermi liquid (FL) when the pseudogap
goes to zero. The backsides instead go into the AFBZ boundary and have weight
exactly equal to zero in this same limit. As seen in the figure, when $x$ moves
towards the Mott insulating state, the Luttinger hole pocket becomes
increasingly short with $\theta_h$ moving toward the nodal direction and
this is the agency whereby the metallicity of the material is increasingly
reduced. The number of states with significant weighting and zero excitation
energy is reduced. The approach to half filling has a progressively stronger
detrimental effect  
on the dynamics of the charge carriers due to the increased magnitude of the 
pseudogap which has opened on the AFBZ boundary.

An important point to note is that the branch $E^-_{\boldsymbol{k}}$ corresponds
to negative energy except for momenta forming the hole pocket, while 
$E^+_{\boldsymbol{k}}$ corresponds to positive energies except for momenta
inside the electron pocket.

\section{Results for the density of quasiparticle states}

\begin{figure}
\includegraphics[width=0.48\textwidth]{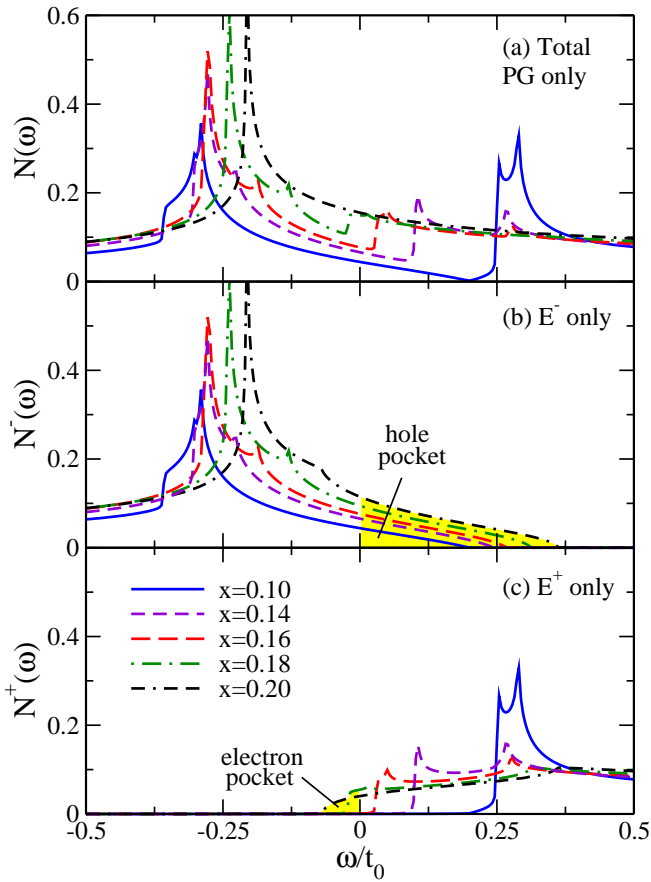}
\caption{\label{fig2} (Color online) Density of quasiparticle states
$N(\omega)$ in the pseudogap state as a function of $\omega$ in units
of $t_0$. (a) shows the total DOS, and (b) and (c) give the partial
results from the $E^-_{\boldsymbol{k}}$ and $E^+_{\boldsymbol{k}}$ branches 
separately.  The shaded yellow region in (b) and (c) identifies the 
contribution of hole and electron pockets, respectively.
}
\end{figure}

In Fig.~\ref{fig2}, we show the quasiparticle DOS corresponding to each band separately
in frames (b) and (c), $N^-(\omega)$ and $N^+(\omega)$, respectively, with
the total $N(\omega)=N^-(\omega)+N^+(\omega)$ given in frame (a). Here,
$N^\pm(\omega)$ is the sum over the Brillouin zone of 
$g_t(x)W^\pm_{\boldsymbol{k}}\delta(\omega-E^\pm_{\boldsymbol{k}})$. The double-dashed dotted black
curve for $N^-(\omega)$ gives results for $x=0.2$ which has no pseudogap and corresponds
to the usual Fermi liquid band structure. If to this we add the contribution
from $N^+(\omega)$ in frame~(c), we obtain
 the usual FL result shown 
as the double-dashed dotted black curve for the total DOS in (a). 
It extends over an energy range
of order several $t_0$ and has a van Hove singularity at an energy slightly
above $-0.25t_0$. The energy scale on which the DOS can vary
significantly, however, is set by the band width except for the rapid variation around
the van Hove singularity, but we will not emphasize this aspect. Returning to the
frames (b) and (c) of Fig.~\ref{fig2}, the areas shaded in yellow for
emphasis correspond, respectively, to the contributions for hole 
and electron  Luttinger pockets. Note in particular that the
shaded region for $N^+(\omega)$ exists only for dopings near optimum
which is the only regime for which
 electron pockets appear. As the doping is decreased 
towards the Mott transition, a gap forms in this band above $\omega=0$
and there is a sharp rise in DOS at some finite frequency above which states
are seen to pile up before $N^+(\omega)$ returns to a value
closer to its no pseudogap value. The energy associated with the sharp rise in
$N^+(\omega)$ can be identified as an effective pseudogap value,
it is different from the input pseudogap although
it is of the same order in magnitude. In fact,
the net effect of the input pseudogap can be a depression which can
even fall at negative energies as can
be seen clearly in the composite curve shown in the top frame~(a)
for $x=0.18$. The DOS
coming from the negative branch  $E^-_{\boldsymbol{k}}$ also shows a clear
effect of pseudogap formation when compared with the double-dashed dotted
 black curve.
We see a shift of the main van Hove singularity (present in the FL case)
to lower energies, followed by  a depression at higher energies beyond a second
relatively smaller van Hove structure. The position in energy of this
second structure  can be taken as a second measure of an effective pseudogap.

\begin{figure}
\includegraphics[width=0.48\textwidth]{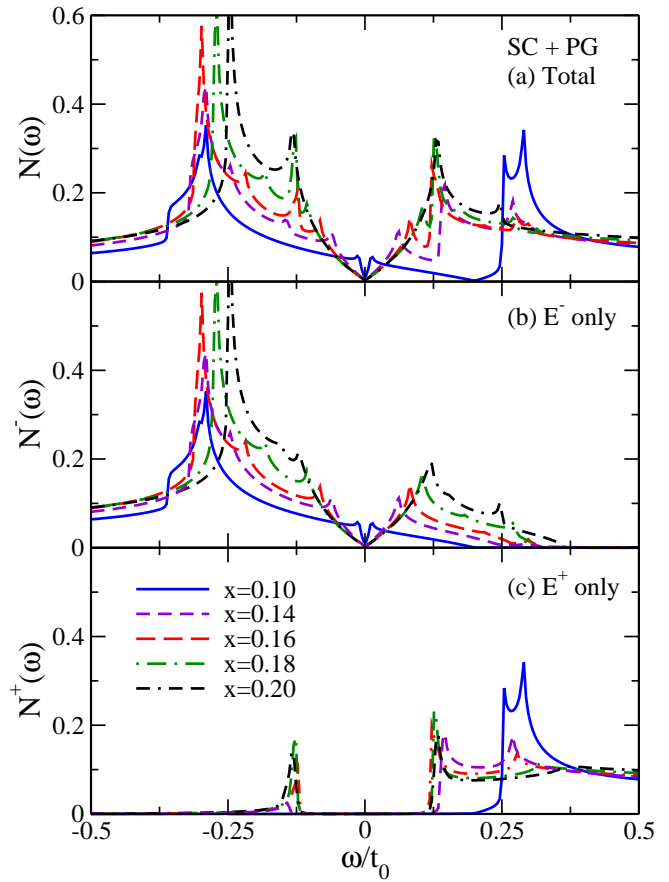}
\caption{\label{fig3} (Color online) Density of quasiparticle states $N(\omega)$
when both the superconducting and pseudogap are included, as a function of
$\omega$ in units of $t_0$. Similar to  Fig.~\ref{fig2}, (a) shows the total
DOS and (b) and (c) give the partial results for $E^-_{\boldsymbol{k},S}$ and $E^+_{\boldsymbol{k},S}$ branches, respectively. 
}
\end{figure}

In the upper curve for the total DOS, the two scales identified as due to the
pseudogap combine to give a dip in the FL DOS which  initially, for small pseudogap,
is confined to the
region of negative energies and is not at the Fermi surface as is often assumed in
phenomenological models of the pseudogap. This dip
grows with decreasing $x$ both in range over which it extends and in depth.
For small $x$ it ranges over a large energy region and becomes a dominant
feature in the DOS which is also greatly reduced around the Fermi energy $\omega=0$.
It is important to emphasize that the pseudogap can introduce significant
anisotropy between positive and negative values of $\omega$ beyond the relatively
mild particle-hole asymmetry of the starting FL band structure. This is an 
essential element of the YRZ model which would not arise if the pseudogap
opened on the Fermi surface rather than on the AFBZ boundary. In this regard
Fig.~\ref{fig3} is particularly relevant. It shows our results when, in addition to a
pseudogap, we include a superconducting gap. Because this second gap opens on
the Fermi surface it produces a new total DOS which is much more particle-hole
symmetric about $\omega=0$ than was the underlying pseudogap-only DOS,
for energies of order of the superconducting gap. Before emphasizing this
important point further, we note that the opening of the superconducting
gap has pushed some spectral weight in $N^+(\omega)$ to lower negative energies
due to the Bogoliubov coherence factors characteristic of Cooper pairing.
The peak at the gap at negative energies is quite significant in magnitude.

\begin{figure}
\includegraphics[width=0.48\textwidth]{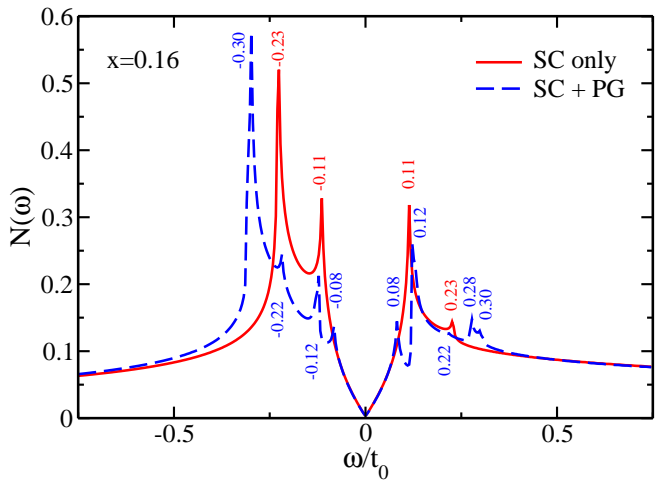}
\caption{\label{fig4} (Color online) Comparison of the total DOS $N(\omega)$
versus $\omega$ in units of $t_0$ for doping $x=0.16$. Superconductivity is 
included in both curves. The dashed blue curve includes the pseudogap.
}
\end{figure}

In Fig.~\ref{fig4}, we compare results of calculations for
$x=0.16$ when only a superconducting gap is present (solid red curve)
and when in addition there is also a pseudogap (dashed blue curve). For the
solid curve, the superconducting coherence peaks fall symmetrically at
$\omega=\pm 0.11$ even though the underlying FL DOS has a van Hove singularity
which appears only at negative bias. There is as well a second peak at
$\omega=\pm 0.23$ which is the normal state van Hove singularity slightly
shifted by the superconductivity and now appearing 
at both positive and negative
biases although its 
positive bias image is greatly suppressed in amplitude. More surprisingly
for the pseudogap case, the symmetry between positive and negative $\omega$
remains in that there are peaks at $\omega=\pm 0.08$ and $\pm 0.12$
as well as at $\omega=\pm 0.22$ and even at $\omega=\pm 0.30$. 
But of course the magnitude of
each peak in a given pair can be very different. The asymmetry prominent
in the nonsuperconducting state has been greatly suppressed by the onset of the
superconducting gap. Finally note that at low bias, the dashed blue curve
and solid red curve agree very well and the introduction of the pseudogap
has had no effect on this region of energy. We will return to this issue 
later in Sec.~V.

\begin{figure}
\includegraphics[width=0.48\textwidth]{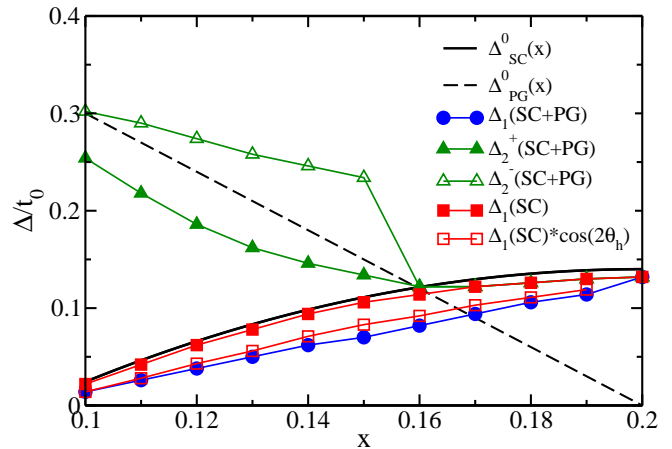}
\caption{\label{fig5} (Color online) Energy scales extracted from 
total quasiparticle density of states as a function of doping $x$.
Black solid and dashed curves are for comparison and are the amplitudes
of the input gap $\Delta^0_{\rm sc}(x)$ and
$\Delta^0_{\rm pg}(x)$, respectively. The solid red squares are taken 
from the superconducting coherence peak position  when the pseudogap
is set to zero and the open red squares are $\cos(2\theta_h)$ times the
value of the solid red squares. With the pseudogap on, the positive 
and negative biases give different energy scales associated with
resolvable peaks in $N(\omega)$. The green triangles 
are the superconducting gap peaks at larger dopings $x\gtrsim 0.16$ and
the solid blue dots are secondary superconducting gap peaks associated
with the end of the Luttinger hole pocket. For $x\lesssim 0.16$, the filled
and open triangles are the pseudogap peaks for positive and
negative bias, respectively. }
\end{figure}

Having just emphasized the restoration of particle-hole symmetry through
superconductivity, we explore next the asymmetry which nevertheless remains
in the quasiparticle DOS and which can be traced to the pseudogap.
In Fig.~\ref{fig5}, the solid black curve gives the input superconducting
gap amplitude $\Delta^0_{\rm sc}(x)$ as a function of doping from $x=0.2$
(the QCP) to $x=0.1$, close to the end of the superconducting dome for the
parameters used by YRZ. The solid red squares were obtained in the FL calculations based on the underlying large Fermi surfaces assuming 
$\Delta^0_{\rm pg}(x)=0$
at all dopings. These points are taken from the energy of the superconducting coherence
peaks in these calculations and fall slightly below the input values for 
$\Delta^0_{\rm sc}(x)$.
This is expected since the peak in the DOS is representative of the extremal
value of the superconducting gap on the Fermi surface rather than in
the Brillouin  zone and these are slightly different.
When both the superconductivity and the pseudogap are present,
the peak structure in $N(\omega)$ becomes more complex. The lowest
energy peaks in the top frame of Fig.~\ref{fig3} are associated
with the superconducting gap. Starting with the most underdoped case first,
solid blue curve for $x=0.1$, we see suppressed coherence peaks as compared
to their magnitude in the optimum case $x=0.2$ (black double dashed-dotted curve).
Also, the energy at which they occur is considerably less than the value of the
input gap amplitude. We plot these in Fig.~\ref{fig5} as the solid blue circles
and see that they trace a dome, but in all cases fall considerably below
the input gap amplitude curve (solid black line). These points correspond to
the gap at the edge of the Luttinger hole pocket and are therefore considerably
smaller in energy than $\Delta^0_{\rm sc}(x)$. In fact, we show as 
open red squares, the product of the solid red squares times $\cos(2\theta_h)$
(see Fig.~\ref{fig1}) and these agree very well with the solid blue circles. In all cases, the 
solid and open green triangles for positive and negative bias, respectively,
are the energies associated with the second significant or
resolvable peak closest to $\omega=0$. These points fall very close to the
superconducting coherence peak energies of the corresponding FL for $x\gtrsim 0.16$ (the triangles are underneath the solid squares for $x\ge 0.17$ in Fig.~\ref{fig5})
and are the same for positive and negative bias $\omega$, {\it i.e.} we have
particle-hole symmetry in this case. We also conclude that for this range of
doping, these peaks are the primary 
superconducting coherence peaks even when the
pseudogap is present and further they remain largely unmodified from the FL
$\Delta^0_{\rm pg}(x)=0$ case. In this sense for these cases, the superconducting
gap has largely overridden the effect of the pseudogap. This all changes for 
$x \lesssim 0.16$. In those cases, the energy of the second peak is quite
different for positive and negative bias and these pseudogap peaks exhibit
a great deal of anisotropy. This second energy scale is also seen in
Raman scattering\cite{sacuto} 
where the nodal $B_{2g}$ geometry probes the superconducting
gap scale and the $B_{1g}$ antinodal geometry, the pseudogap.  As we have
seen here, when optimum doping is approached, the two scales are found to
merge into a single superconducting gap scale. What is different, however,
is that the DOS peak structure can be exploited to get information on the
asymmetric effect of the pseudogap between positive and negative
energies. Our results in Fig.~\ref{fig5} show that this can be considerable
and that the effect sets in quite abruptly around $x=0.16$ when only
the hole Luttinger pocket remains.

\begin{figure}
\includegraphics[width=0.48\textwidth]{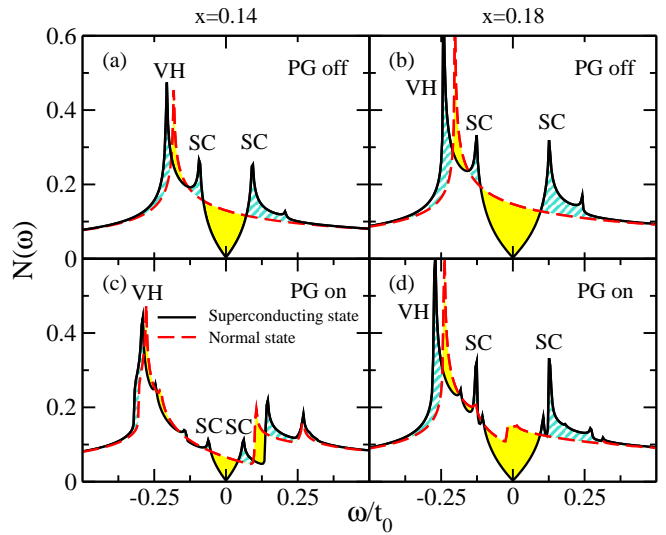}
\caption{\label{fig6} (Color online) 
The quasiparticle DOS $N(\omega)$ with the pseudogap off [(a) and (b)] 
and on [(c) and (d)] for two doping values $x=0.14$ (left frames) and 
0.18 (right frames). Comparison is made between the normal state
(red dashed curve) which has no superconductivity and the case with
superconductivity (black solid curve). The yellow solid shading and
the blue hatched shading help to see the conservation of states.
}
\end{figure}

Fig.~\ref{fig6} shows results for the DOS at two dopings $x=0.14$ (left frames)
and $x=0.18$ (right frames). The top frames compare results for normal (red
dashed curve) and superconducting state (black solid curve). The yellow solid
shading and blue hatched regions help one to see that our calculations
 conserve the number of states between the normal and
superconducting case as they must. 
What is clear from the figure and what we wish to emphasize 
here is that when there is no pseudogap, the states lost below the 
superconducting gap are largely recovered just above the coherence peak
except for a small shift in the van Hove (VH) singularity which introduces
a slight complication that would not be present in models with a constant
DOS. When the pseudogap is included there is further readjustment
of spectral weight introduced by the transition to the superconducting state.
The energy range over which there remains
significant changes is now set by 
the pseudogap scale. Here we are not emphasizing
the slight complication brought about by the existence of a van Hove
singularity in the underlying band structure chosen in the YRZ model.

\section{Decomposition of DOS in angles in the Brillouin zone}

\begin{figure}
\includegraphics[width=0.48\textwidth]{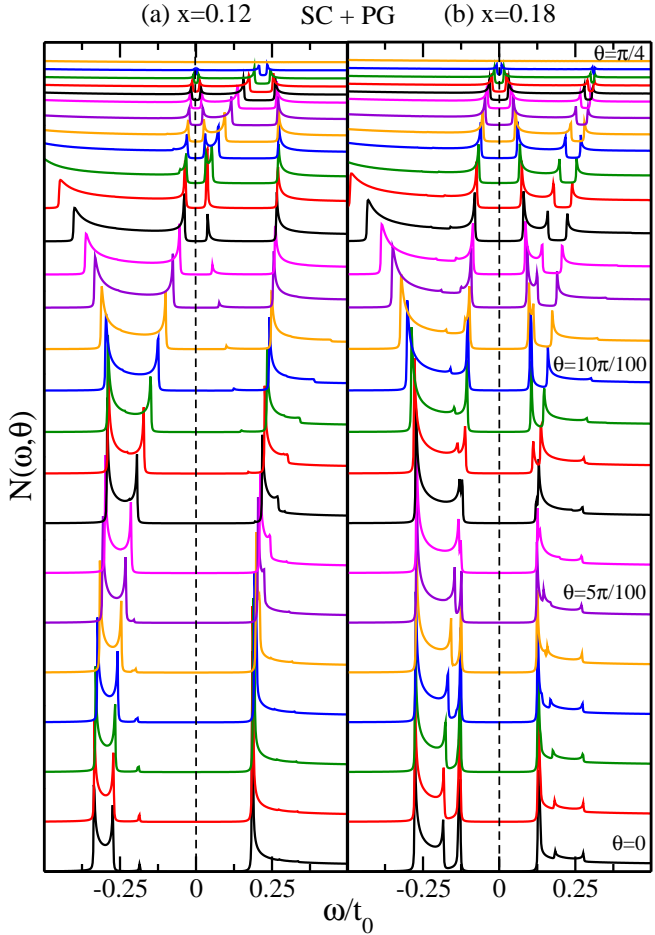}
\caption{\label{fig7} (Color online) The partial DOS $N(\omega,\theta)$
as a function of $\omega$ in units of $t_0$ for two values of doping
(a) $x=0.12$ and (b) $x=0.18$. From top to bottom, the curves are
shown from $\theta=\pi/4$ to $\theta=0$ in steps of $\pi/100$.
}
\end{figure}

We turn next to the decomposition of the total DOS into partial
contributions coming from different angular slices in the Brillouin
zone. Such a decomposition has been considered by Pushp et al.\cite{pushp}
in relation to their experimental scanning tunneling spectroscopy (STS)
results. They effectively decompose the total $N(\omega)$ by writing
\begin{equation}
N(\omega)\equiv\int_0^{2\pi}\frac{d\theta}{2\pi}N(\omega,\theta),
\label{eq:Nint}
\end{equation}
where for $N(\omega,\theta)$, they take the explicit analytic
functional form
\begin{equation}
N(\omega,\theta)=Re\biggl\{\frac{\omega-i\Gamma}
{\sqrt{(\omega-i\Gamma)^2-\Delta(\theta)^2}}\biggr\}W(\theta),
\label{eq:Neff}
\end{equation}
where $\Gamma$ is a smearing parameter to be varied to get a best fit of
the data along with the value of the gap $\Delta(\theta)$
and weighting factor $W(\theta)$. In this
way, a gap scale $\Delta(\theta)$ can be obtained as a function of
$\theta$. Inspired by the above, we decompose the full density
of states
\begin{equation}
N(\omega)=\sum_{\boldsymbol{k}}A(\boldsymbol{k},\omega),
\end{equation}
where $A(\boldsymbol{k},\omega)$ is given in Eq.~(\ref{eq:A}), into an integration over the magnitude of momentum $k$
for fixed  angle $\theta$ in the Brillouin zone. This gives 
$N(\omega,\theta)$ directly and results are presented in Fig.~\ref{fig7}.
Frame (a) is for doping $x=0.12$ and (b) for $x=0.18$. Twenty-six values
of $\theta$ are shown between $45^\circ$ and $0^\circ$ as measured from
the $(\pi,\pi)$ point of the Brillouin zone (see Fig.~\ref{fig1}).
The nodal direction (top curves) show small coherence peaks due to superconductivity
which are 
centered about $\omega=0$ and are particle-hole symmetric. At higher positive 
energies, we see that there are two more pseudogap peaks while at negative
bias, a van Hove singularity is seen.  

\begin{figure}
\includegraphics[width=0.48\textwidth]{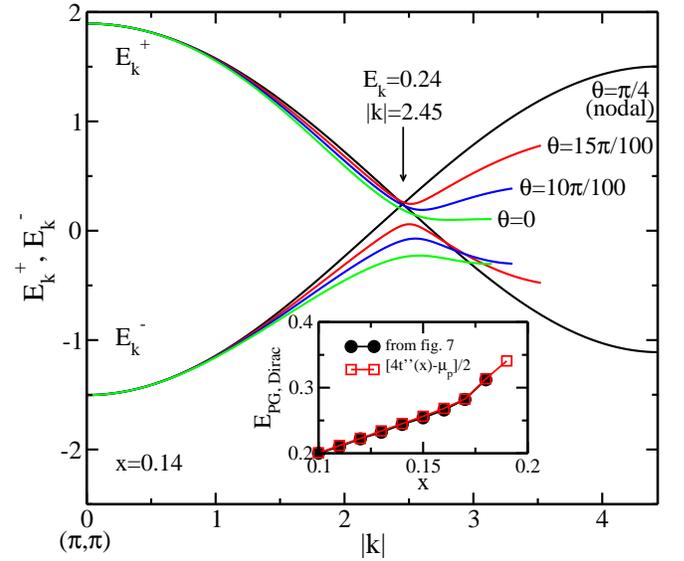}
\caption{\label{fig8} (Color online) 
The variation of the energies $E^+_{\boldsymbol{k}}$ and $E^-_{\boldsymbol{k}}$ with magnitude of momentum $k=|\boldsymbol{k}|$,
measured from ($\pi,\pi$),  for various fixed values of the angle $\theta$
defined in Fig.~\ref{fig1} as indicated in the figure. The inset gives
the energy of the pseudogap Dirac point as a function of doping $x$.
Results for our complete numerical calculations  from curves
as shown in Fig~\ref{fig7} are compared with
the approximate formula $E_{\rm dirac}=(4t''-\mu_p)/2$.
}
\end{figure}

In Fig.~\ref{fig8}, we show results
for $E^+_{\boldsymbol{k}}$ and $E^-_{\boldsymbol{k}}$ 
as a function of absolute value of 
momentum $|\boldsymbol{k}|=k$ measured from $(\pi,\pi)$ for
several values of angle $\theta$ measured similarly as the angles
shown in Fig.~\ref{fig1} and used for Fig.~\ref{fig7}.
 Doping was set at $x=0.14$. The pseudogap
Dirac point falls at $\theta=\pi/4$ and $k=2.45$ in units of 
$1/a$ (solid black curves) and this corresponds to
$E^+_{\boldsymbol{k}}=E^-_{\boldsymbol{k}}=0.24$ in units of the unrenormalized
 nearest neighbor hopping parameter $t_0$. As the angle $\theta$ is increased
away from the nodal direction, the energies
$E^+_{\boldsymbol{k}}$ and $E^-_{\boldsymbol{k}}$  no longer
meet but split and there is a clear gap between them. Both move
down in energy but not by the same amount. Also, the maximum and minimum
corresponding to a given pair of curves do not fall at exactly the same
value of $k$. In the above sense, the two van Hove singularities at
issue which define the pseudogap peaks in $N(\theta,\omega)$ of Fig.~\ref{fig7}
are not entirely symmetric. The energies of the Dirac point can easily be 
traced as a function of doping. It corresponds to the point 
$E^+_{\boldsymbol{k}}=E^-_{\boldsymbol{k}}$  which occurs for 
$\xi_{\boldsymbol{k}}+\xi^0_{\boldsymbol{k}}=0$, {\it i.e.} 
$E_{\boldsymbol{k}}=\sqrt{\tilde\xi^2_{\boldsymbol{k}}
+\Delta_{\rm pg}^2(\boldsymbol{k})}=0$ for $\theta=\pi/4$,
and the details of the dispersion curves determine the critical
value of $k\equiv k_c$. For $x=0.14$, this $k_c=2.45$
in Fig.~\ref{fig8} which is somewhat larger than the value
$\pi/\sqrt{2}$ which corresponds to $k_x=k_y=\pi/2$. For this latter
point, the dispersion curves of YRZ are particularly simple and only third
neighbor hopping survives. Taking $E_{\boldsymbol{k}}=0$ but using
$k_x=k_y=\pi/2$ to evaluate the remaining expression,
provides an approximate estimate for the energy
of the Dirac  point  $E_{\rm dirac}$ as $[4t''(x)-\mu_p]/2$. This estimate is
shown in the inset of Fig.~\ref{fig8} to be quite adequate and shows how
$E_{\rm dirac}$ moves towards zero energy as $x$ decreases, as is
also seen in Fig.~\ref{fig7} where the pseudogap Dirac
point is seen as the closing of the pseudogap coherence peaks at $\theta=\pi/4$. 

\begin{figure}
\includegraphics[width=0.48\textwidth]{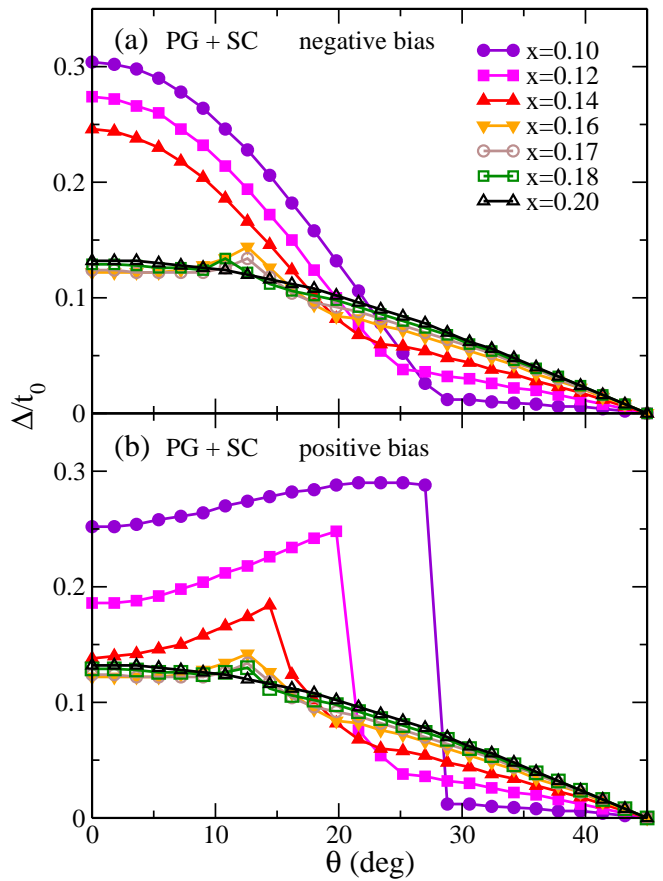}
\caption{\label{fig9} (Color online) 
The energy gap read off the curves for $N(\omega,\theta)$ of the type
shown in 
Fig.~\ref{fig7} as a function of $\theta$ in the Brillouin zone measured
from $(\pi,\pi)$. The top frame is for negative bias (occupied states)
and the bottom for positive bias (unoccupied states). Note the isotropy
between these two sets of data for angles around 45$^\circ$ or
$\pi/4$ (nodal direction)
and the large anisotropy at small angles (antinodal).
}
\end{figure}

Returning now to this Fig.~\ref{fig7}, we note that
as $\theta$ is
decreased
towards the antinodal direction, it remains possible to identify 
unambiguously a lowest energy peak and the magnitude of the
energy at which these peaks
occur is recorded on Fig.~\ref{fig9} as a gap in units of $t_0$ for each
angle. For optimum or even near optimum doping, the resulting curve is close to
a simple $\cos(2\theta)$ curve and represents the superconducting gap. However,
it should be noted that the curves for $x=0.18$, 0.17 and 0.16 all show
an additional structure between $10^\circ$ and $15^\circ$ and the
overall curve does deviate visibly from a simple $\cos(2\theta)$.
This is not surprising since near $x\lesssim 0.2$, the Fermi surface, as we have
shown in Fig.~\ref{fig1}, has reconstructed from the large open surface of
FLT to Luttinger surfaces (holes about the nodal region and electrons about
the antinodal point) and the partial DOS $N(\omega,\theta)$ has some 
knowledge of this fact. Consequently, the extracted energy scale is
not a pure superconducting gap scale. Nevertheless, the distortions from
a pure $\cos(2\theta)$ are not large and the curves show almost perfect
particle-hole symmetry (compare the top and bottom frame of Fig.~\ref{fig9} for
$x\geq 0.16$). 

The situation is quite different and more interesting for low dopings approaching
closer to the Mott transition. In that case, only a Luttinger hole pocket remains as shown
in Fig.~\ref{fig1}. When superconductivity is not considered, there are
zero energy excitations along the Luttinger contours which are real Fermi surfaces
and these are gapped by superconductivity. But for angles smaller than
$\theta_h$, no true  Fermi surface exists and consequently, the peaks
seen in $N(\omega,\theta)$ [Fig.~\ref{fig7}(a)] no longer have their origin
in $\Delta_{\rm sc}(\boldsymbol{k})$ but are related rather to the pseudogap
$\Delta_{\rm pg}(\boldsymbol{k})$. This can be easily traced in Fig.~\ref{fig7}(a)
for $x=0.12$. In the nodal direction, the peaks nearest $\omega=0$ are
the superconducting coherence peaks and these become more prominent as the
gap opens up. But eventually, particularly on the positive bias side,
they start to lose intensity while at the same time, the pseudogap peak
at higher energy remains quite intense and it is this peak that must eventually
be taken if we are to characterize $N(\omega,\theta)$ for smaller
values of $\theta$ with a single energy scale as we are doing here. 
It is clear that we need to jump from one energy 
scale to the
other. On the negative bias side, however, the progression with decreasing
$\theta$ remains smooth. We always take the large intensity peak closest to the
origin $\omega=0$. These facts are reflected in our Fig.~\ref{fig9}. In the
top frame for $x\leq 0.14$, we see two very distinct gap scales: a 
superconducting one for $\theta$ near $45^\circ$ and a pseudogap one for $\theta$
going towards zero. The curves have a kink at an angle corresponding to
$\theta=\theta_h$, the end of the Luttinger pocket, but otherwise they show a 
rather smooth behavior. On the other hand, for positive biases (lower frame
of Fig.~\ref{fig9}) there is a clear jump in the curves as we transfer
from superconducting to pseudogap scale and this is followed by a 
progressive drop to lower values as $\theta$ decreases towards zero
in sharp contrast to the top frame for negative $\omega$.

\begin{figure}
\includegraphics[width=0.48\textwidth]{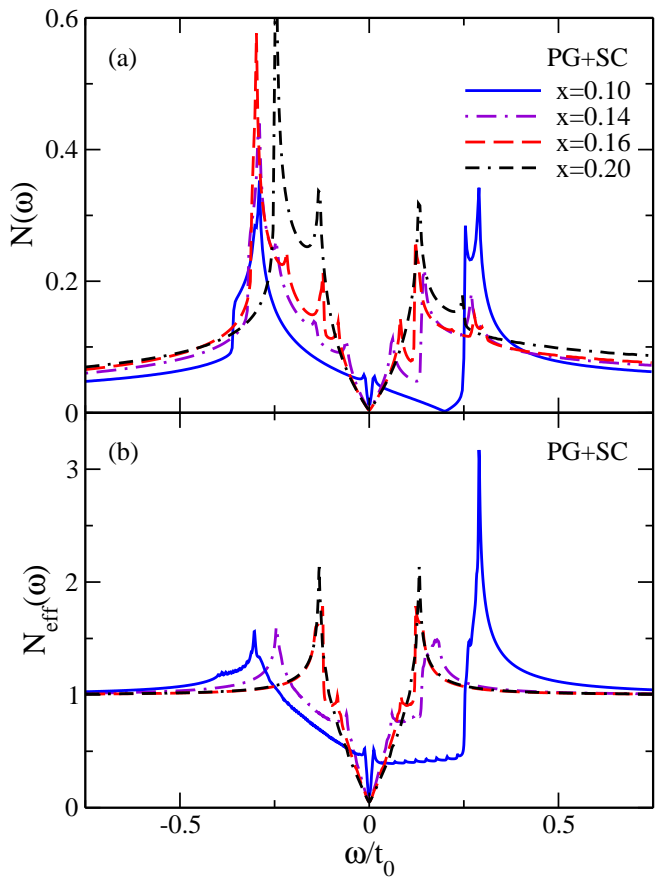}
\caption{\label{fig10} (Color online) 
Comparison of (a) the original DOS $N(\omega)$ from Fig.~\ref{fig3}
with (b) the  quasiparticle DOS $N_{\rm eff}(\omega)$ 
versus $\omega$ in units of $t_0$ as defined by Eqs.~(\ref{eq:Nint})-(\ref{eq:Neff}). 
Both (a) and (b) refer to the state
with superconductivity and the pseudogap present. 
}
\end{figure}

To make clearer that our results for the frequency
dependence of $N(\omega,\theta)$ cannot be fit by the simple formula
of Eq.~(\ref{eq:Neff}), we have used this formula
along with the gaps presented in Fig.~\ref{fig9} to recalculate a total
DOS $N(\omega)$ according to Eq.~(\ref{eq:Nint}) with
weights $W(\theta)$ set equal to one for simplicity. Numerical results
are presented in Fig.~\ref{fig10}(b). For ease of comparison, we have
reproduced in Fig.~\ref{fig10}(a) 
the full DOS already presented in Fig.~\ref{fig3} 
on which our angular decomposition is based. 
For small values of $\omega$, we see a great deal of agreement between
these two sets of figures. Near optimum doping only the superconducting
gap scale is prominent while for the highly underdoped regime both
superconducting and pseudogap scales are clearly seen. At larger energies
important differences arise and these have their origin in the failure of
Eq.~(\ref{eq:Neff}) with a single energy scale to capture all the
details present in the partial densities of states of Fig.~\ref{fig7}.
In particular, the van Hove singularity seen on the negative bias side in the
top frame is completely missed in the lower frame. While this figure 
speaks to the limitation in the method used to extract energy gaps
from STS data\cite{pushp} it also shows
that important  qualitative and even semiquantitative information can be 
extracted in this way. Some details are certainly missed but other important
features are quite prominently seen such as the pseudogap scale and its
asymmetry. One could improve the agreement between the top and bottom
frame in Fig.~\ref{fig10} by including weights $W(\theta)$ or multiplying
$N_{\rm eff}(\omega)$ by an envelope function, which partially
accounts for the presence of the van Hove singularity present
in our model band structure, but this is not our aim here.

\begin{figure}
\includegraphics[width=0.48\textwidth]{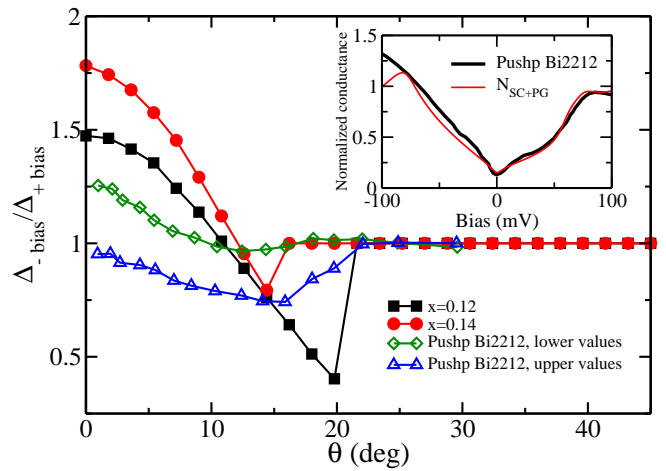}
\caption{\label{fig11} (Color online) 
The ratio of $\Delta_{-\rm bias}/\Delta_{+\rm bias}$ versus angle,
taken from Fig.~\ref{fig9} for $x=0.12$ and 0.14. The theoretical work
is compared with the data of Pushp et al.\cite{pushp} which is bounded by
the diamonds and triangles (see text for details). Inset:
a comparison between the theory and experiment for the normalized conductance
(thick black curve) of Pushp et al. and a DOS curve from our calculations
(thin red curve). }
\end{figure}

In Fig.~\ref{fig11}, we provide a direct comparison between our results
for the asymmetry between positive and negative bias of the derived angular
dependent gap and the data of Pushp et al.\cite{pushp} for the same 
quantity. Plotted is the ratio of $\Delta_{-\rm bias}/\Delta_{+\rm bias}$
from our Fig.~\ref{fig9} for $x=0.12$ (squares) and 0.14 (circles). Pushp
et al.\cite{pushp} show several curves in their figure S3 for 
a range of data taken from several spots on their 
UD58 sample. As the negative bias data is fairly uniform, we took points from 
along the trend of the
data. For the positive bias, there is a range of values
for each angle and so we took the highest and lowest values 
to form separately the ratio with the negative bias data. This is
shown as the blue 
triangles and green diamonds, respectively.
As we wish to concentrate on the asymmetry, only the data for
angles less than 30$^\circ$ are shown. Due to the expected
symmetry for angles in the nodal region, the data ratio towards
the nodal region have been normalized to one to facilitate 
comparison with theory
even though the ratio in the data was slightly greater than one. Our procedure is
not entirely rigorous and in the
hands of the experimentalists there might be a more accurate analysis
of the result. Nevertheless, while the individual curves for $\Delta_+$
and $\Delta_-$ are quite different in shape as compared with our theoretical
results, we find that
the ratio shows the same qualitative trend as theory. There  is a
significant dip around 15-20$^\circ$ followed by a rise as
the antinodal region is approached. This comparison provides evidence that the
pseudogap forms asymmetrically about the Fermi surface as in our model.
In the inset to Fig.~\ref{fig11}, we show results for the DOS in
the resonating valence bond spin liquid compared with UD35 data from
Pushp et al.\cite{pushp} (their figure S4). We used
$x=0.11$ to match the $T_c$ reduction from optimum and have added broadening
and a linear
incoherent background to provide a better fit to the data. This linear
background does not alter the structures due to the two energy
gaps in the model. The agreement is good in the low energy region shown and it
again clearly reveals asymmetry between positive and negative biases.

\begin{figure}
\includegraphics[width=0.48\textwidth]{fig12a.eps}
\includegraphics[width=0.48\textwidth]{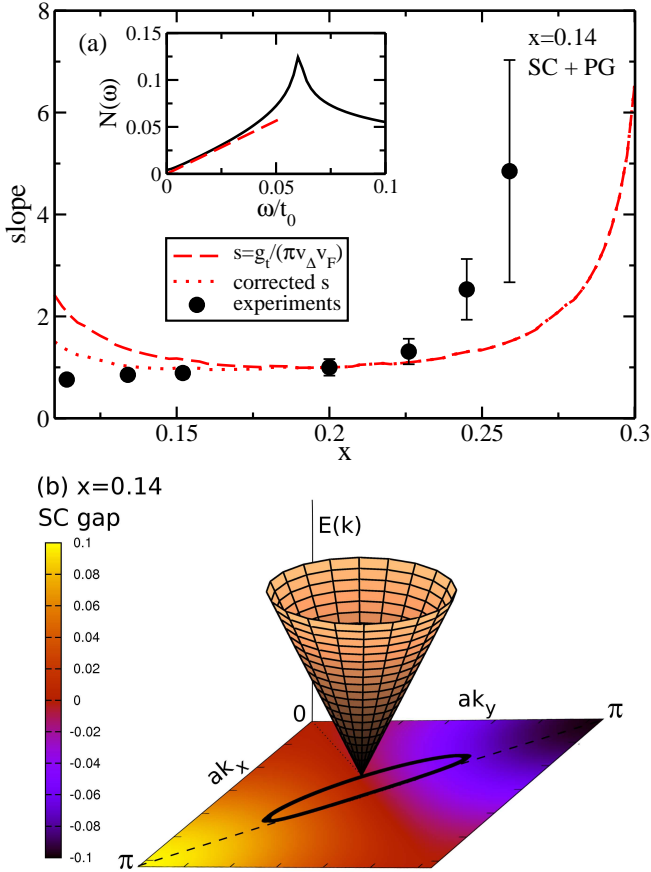}
\caption{\label{fig12} (Color online) (a) The slope
of $N(\omega)$ normalized to that for optimal doping versus $x$
compared with the data from Pushp {\it et al.}\cite{pushp}.
Inset: $N(\omega)$ for $x=0.14$ with both
the superconducting gap and pseudogap present showing the linear behavior at
low $\omega$ in agreement with Eq.~(\ref{eq:slope}) which 
is shown as the red dashed line.
 (b) A schematic of 
the superconducting
 Dirac cone as it grows in energy out of the nodal region on the 
heavily weighted side of the Luttinger hole
pocket. The heavy solid black curve is the Luttinger contour for $x=0.14$.
}
\end{figure}

\section{Zero frequency limit}

We finally turn to the slope of the DOS at $\omega\to 0$ shown in
Fig.~\ref{fig12}(a). In YRZ theory, the
superconducting
Dirac point [shown schematically by Fig.~\ref{fig12}(b)],
on the heavily weighted side of the Luttinger hole pocket in 
the nodal direction, is not affected by pseudogap formation. For
small energies $\omega$, it is clear from the cone shown in
Fig.~\ref{fig12}, that the only excited states available  are those
near the bottom tip of the Dirac cone.  We can
easily show that, as for an ordinary FL\cite{durst}, the slope is given by
\begin{equation}
s=\frac{g_t}{\pi v_\Delta v_F}
\label{eq:slope}
\end{equation}
with $N(\omega)=s|\omega|$. Here, $v_F$ is the Fermi velocity
at the Dirac point and 
$v_\Delta=|\nabla\Delta_{\rm sc}(k)|_{k_F}=(\Delta^0_{\rm
  sc}/\sqrt{2})|\sin(k_{Fx})|$,
is the superconducting gap velocity
at this same point. In the inset of Fig.~\ref{fig12}(a), we show results of
complete numerical calculations of the DOS for the case of $x=0.14$
with both pseudogap and superconducting gap included (solid black curve)
and compare with the results of Eq.~(\ref{eq:slope}) (dashed red line).
We see good correspondence.

Formula (\ref{eq:slope}) is the same as would hold in
a FL with two very important differences: $g_t$ and a possible
variation of $v_\Delta$ in this model. In Eq.~(\ref{eq:slope}), 
the Gutzwiller coherence factor
$g_t(x)$  carries the information on how
much weight remains in the coherent part of the Green's function when
correlation effects are included, with the rest shifted to incoherent
processes. This factor is a crucial part of our present approach but does not
enter FLT. An important consequence of this fact is that it 
reduces the rise in the slope in the highly underdoped regime
of the cuprate phase diagram as compared with a pure FL
approach. It provides a factor of $2x/(1+x)$ while the
gap velocity is proportional to $T_c$. Pushp et al.\cite{pushp} 
found in STS data that the slope
 was nearly constant over a significant doping range below optimum.
Matching their reduction from optimum of the $T_c$ of their samples with
our superconducting dome in order to determine the relation to
the doping $x$ used in our
model, we plot the slope of their data normalized to the value at optimum
versus $x$. As there was a range of values in their data for a particular doping, 
we have done our best to indicate this as a point with a bracketing bar.
Comparison with this data (solid black circles in Fig.~\ref{fig12})
 shows that, around $x\simeq 0.10$ which is the
doping in our calculations that corresponds to a reduction of $T_c$ by a factor
of 3
below its value at optimum, the predicted slope represented by
the dashed red curve starts to increase while
experiment does not. This could mean that in reality the Gutzwiller factor
is a more strongly decaying function of $x$ than the one we have used.
Alternatively, broadening will have the tendency to decrease the slope.
However, so far we are neglecting another important effect associated with
the YRZ model. In this model
 the gap to $T_c$ ratio 
$2\Delta^0_{\rm sc}/k_BT_c$, which for simplicity we have fixed at a value of
6 in all our calculations, 
is known to vary importantly with $x$. In very recent work, Schachinger
and Carbotte\cite{ewaldyrz} have solved a  generalized BCS gap equation
with the pseudogap and Fermi surface reconstruction fully accounted for
and have found that this ratio changes from its canonical value of $4.3$
at $x=0.2$ (optimum) to $\sim 6.5$ or even higher 
towards the end of the dome as the Mott insulator
and antiferromagnetism is approached. Accounting for this reduces
the variation in slope between optimum and highly underdoped by $\sim 50$\%,
bringing our calculations much closer to experiment as indicated by the
red dotted curve in Fig.~\ref{fig12}(a). The overall agreement between theory
and the data is very good.
It is important to stress that $g_t(x)$ provides an 
important factor in Eq.~(\ref{eq:slope}) which brings theory much
closer to experiment than what one would find in a FL approach
and this also is true as well for the variation of gap to critical
temperature ratio.

\section{Summary and Conclusions}

We have computed the total quasiparticle DOS in the resonating valence bond
spin liquid 
model\cite{yrz}
 of the underdoped cuprates. When superconductivity is not included,
the formation of the pseudogap, which provides a mechanism for Fermi
surface reconstruction, modifies the underlying Fermi liquid DOS in an
asymmetric way with respect to the Fermi energy $(\omega=0)$. For small
values of the pseudogap just below the critical doping associated
with a QCP, the resulting depression in $N(\omega)$ is confined to
negative energies. As doping is reduced  towards the Mott
insulating state, the depression deepens, covers a larger range in
energy and spans positive as well as negative biases but its effect
remains asymmetric. However, if superconductivity is also included,
particle-hole symmetry is restored at lower energies of order 
$\Delta_{\rm sc}^0(x)$, although beyond this range the asymmetry associated
with the pseudogap remains. This effect is traced to the fact that
the pseudogap is associated with the antiferromagnetic Brillouin
zone rather than the Fermi surface where the superconducting gap opens.

One can trace peaks in the total quasiparticle DOS which are,
at optimum doping and just below, coherence peaks due to superconductivity
but these evolve for $x\lesssim 0.16$ into pseudogap peaks which are
distinctly different for positive and negative biases. This asymmetry
is an intrinsic part of the YRZ model and can be used to test its
validity. We are also able to trace a second set of peaks associated with the
superconducting gap which however originate from  the end of the
Luttinger hole pocket at an angle $\theta_h$ in the Brillouin zone.
In the heavily underdoped region of the phase diagram, these peaks are the only
ones that can be identified as due purely to the superconducting gap.
These signatures are relatively weak, however, as found in, for example, the experiments
of Boyer et al.\cite{boyer}, in comparison with those seen in an
ordinary $d$-wave BCS superconductor. 
As a function of doping, the energy 
corresponding to these
coherence structures follow the dome associated with the critical temperature
although the scale of the effective gap involved is reduced. 

When the total DOS is decomposed into partial contributions from each angle 
$\theta$ in the Brillouin zone separately, $N(\omega,\theta)$, it is
found that at some angles $N(\omega,\theta)$ can show complex behavior
as a function of energy $\omega$  and even display several peaks. Such
distributions cannot be modeled by the simple function of Eq.~(\ref{eq:Neff}).
Nonetheless, we find that we could identify,
in a fairly unambiguous fashion, a single peak in these distributions
which was both close to the origin $\omega=0$ and had significant
amplitude.
This provides a single energy scale for each direction $\theta$.
At angles near the nodal direction, this scale is associated 
with the superconducting gap but as
$\theta$ is decreased it progresses into a peak associated with the
pseudogap energy. In some cases the transition from one scale to the other
is smooth while in others it can be abrupt and also somewhat
ambiguous. What
is found is that the superconducting gap peak at low energies rapidly
loses intensity while the pseudogap peak at higher energies is quite intense
and so dominates over the lower energy structure
and thus must be chosen as characteristic of a single gap scale for
this angle $\theta$ in the Brillouin zone. This gap
can however differ considerably in size depending on the sign of the bias
involved. For energy below the superconducting gap scale, there is little
asymmetry but this changes radically as the antinodal direction is approached
in the highly underdoped regime where the pseudogap is probed. Our findings
have implications for the analysis of STS data when one wishes to extract
directional information from such experiments.\cite{pushp}
Comparison with the data of Pushp et al. for both $N(\omega)$ and 
$\Delta_{-\rm bias}(\theta)/\Delta_{+\rm bias}(\theta)$ shows indeed an asymmetry in the DOS
which is in qualitative agreement with the YRZ model.

Another result is that the zero energy slope of the DOS is importantly
reduced from its FL value in YRZ theory because of the appearance of
a Gutzwiller factor $g_t(x)$ which is a rapidly decreasing function as
$x$ decreases and represents the reduction, due to correlation effects,
of the coherent part of the charge carrier Green's function. This factor
can partially compensate for the rapid increase in slope that would occur
 due to the appearance
of the gap velocity $v_\Delta$ in the denominator of the expression for the
slope in ordinary BCS theory. 
However, this factor is modulated by a rapid increase in the value of
gap to critical temperature found to be the direct consequence of an
increase in the pseudogap in the theoretical work of Schachinger and
Carbotte.\cite{ewaldyrz}
Without these two effects
there would be a serious conflict between theory and the experimental
results of Pushp et al.\cite{pushp} who find a slope which remains
fairly constant in the underdoped regime. While a comparison with the
existing data and YRZ predictions gives qualitative agreement,
there are indications that the Gutzwiller factor $g_t(x)$ 
and/or the gap to $T_c$ ratio may in fact
vary more rapidly with decreasing $x$ than indicated by theory.

\begin{acknowledgments}
We thank James LeBlanc for his assistance and helpful discussions.
This work has been supported by NSERC of Canada 
and by the Canadian Institute for Advanced Research (CIFAR).
\end{acknowledgments}

\end{document}